\documentclass[12pt]{article}
\usepackage{epsf}
\usepackage{psfig}
\usepackage{latexsym}
\newcommand{\lsim}{\mathrel{\lower4pt\hbox{$\sim$}}
\hskip-12.5pt\raise1.6pt\hbox{$<$}\;}

\newcommand{\gsim}{\mathrel{\lower4pt\hbox{$\sim$}}
\hskip-12.5pt\raise1.6pt\hbox{$>$}\;}

\newcommand{\be}{\begin{equation}}
\newcommand{\ee}{\end{equation}}
\newcommand{\bea}{\begin{eqnarray}}
\newcommand{\eea}{\end{eqnarray}}
\newcommand{\noi}{\noindent}
\newcommand{\epe}{\epsilon^\prime/\epsilon}

\begin{document}
\title{KAON 99 --- Summary and Perspective}

\author{William J. Marciano \\
Brookhaven National Laboratory \\ Upton, NY\ \ 11973 \\ }

\maketitle

\begin{abstract}
An overview of KAON 99 with commentary is presented. Emphasis is placed
on the state of CKM mixing and CP violation. The Jarlskog invariant,
$J_{CP}$, is shown to provide a useful quantitative comparison of $K$ and $B$
phenomenology. The potential of future rare and ``forbidden'' decay
experiments to probe ${\cal O}$(3000 TeV) ``New Physics'' is also described.
\end{abstract}

\section{Conference Overview and Commentary}

For more than 50 years, Kaon physics has played a leading role in
unveiling Nature's fundamental intricacies and challenging our creative
imaginations \cite{dalitz,wolf}. The 
concept of hadronic ``flavor'' has its roots in the associated
production of kaons and introduction of ``strangeness'' as a nearly
conserved quantum number. SU(3)$_F$, current algebra, and the quark
model all stemmed, to a large extent, from extensive follow-up studies
of that discovery.

The $\theta$-$\tau$ puzzle in $K\to 2\pi$ and $3\pi$ (final states with
different parities) provided the stimulus for Lee and Yang's parity
violation conjecture. Today, we easily accommodate parity violation via the
chiral nature of the Standard Model's SU(2)$_L\times$U(1)$_Y$ local
gauge symmetry. However, that left-right asymmetry remains a deep
fundamental mystery with potentially profound implications about the
short distance properties of space-time and origin of mass.

Early null results in rare $K$ decay searches  also led to 
important physics insights. The observed suppression of flavor changing
neutral currents (FCNC) in $K_L\to\mu^+\mu^-$, $K\to \pi\nu\bar\nu$
etc.\ motivated the G.I.M. (Glashow-Iliopoluous-Maiani) mechanism and
introduction of charm. Today, medium rare ${\cal O}(10^{-8})$, branching
ratios such as $K_L\to \mu^+\mu^-$ are routinely measured with high
precision and used to search for or constrain potential ``New Physics''
effects. 

The special (unique) $\Delta S=2$ mixing features of the $K^0$-$\bar
K^0$ system allowed CP violation to be unveiled in $K_L\to 2\pi$
decays. To explain that enigmatic effect, Kobayashi and Maskawa (KM)
boldly proposed \cite{koba} the now discovered third generation of quarks, $t$
and $b$. Their parametrization of CP violation via angles and phases in
a unitary $3\times3$ quark mixing matrix provided a simple but elegant solution
to that outstanding puzzle. It also suggested many interesting
predictions for FCNC and direct CP violation effects \cite{gilman}. The
recent measurement of $\epsilon^\prime/\epsilon$ in $K\to 2\pi$ and
initial studies of $B\to J/\psi K_s$ lend strong support to their
hypothesis. 

In a sense, the KM model of CP violation trivialized that previously
mysterious phenomenon. It suggested that a mere non-vanishing  weak
interaction phase and quark mixing, rather than some new superweak
interaction was 
responsible for CP violation. That beautiful solution now seems almost
obvious. Also, if 
additional new interactions are eventually uncovered, it seems likely
that they will similarly  have relative phases which would provide additional
sources of CP violation. That would be a welcome discovery, since
electroweak baryogenesis \cite{worah} seems to require additional CP violation beyond the
Standard Model.

Given its already rich and glorious history, what more can we hope to
learn from $K$ decays? Are kaon studies passe, or  competitive with $B$
physics and
other ways to investigate CP violation?

This conference is proof of the excitement $K$ physics continues to
generate. Its copious production cross-section and relative long
lifetime  make the kaon very special and experimentally
popular worldwide. Indeed, there are many ongoing diverse experimental
programs at labs around the world along with exciting ideas and proposals for
new initiatives. I give in Table~\ref{tab1} a list of the kaon programs
discussed at this meeting. Experiments at those facilities measure CKM
(Cabibbo, Kobayashi, Maskawa)
matrix elements, probe CP and possible CPT violation, search for very
rare or forbidden decays, etc. In addition, they thoroughly study
medium rare decays and other properties of kaons, thus providing an
arena for refining theoretical skills such as chiral perturbation
theory, lattice techniques, large $N_c$ approaches and perturbative QCD\null.

\begin{table}[htb]
\begin{center}
\caption{Ongoing and future kaon physics programs reported on at this
meeting \label{tab1}}
\begin{tabular}{ll}
KEK 12 GeV PS${}\to 50$ GeV PS & (talks by T. Inagaki, G.-Y. Lim) \\
BNL 30 GeV AGS + Booster & (talks by L. Littenberg, W. Molzon, \\
& \qquad M. Zeller) \\
FNAL $K$TEV${}\to{}$KAMI & (talks by P. Cooper, J. Whitmore) \\
CERN SPS & (talks by G. Kalmus, L. Koepke) \\
CP LEAR (Completed) & (talk by P. Bloch) \\
Frascati-DA$\phi$NE & (talks by P. Franzini, S. Di Falco) \\
Novosibirsk & (talks by L. Landsberg, N. Ryskulov) 
\end{tabular}
\end{center}
\end{table}

With regard to determining the CKM quark mixing matrix, $V_{\rm CKM}$,
$K$ and $B$ measurements both play special key roles. Their importance is well
illustrated by the Wolfenstein parametrization \cite{wolftwo}

{\footnotesize\be
V_{\rm CKM} = \left(\begin{array}{ccc}
V_{ud} & V_{us} & V_{ub} \\
V_{cd} & V_{cs} & V_{cb} \\
V_{td} & V_{ts} & V_{tb} 
\end{array}\right) =
\left( \begin{array}{ccc}
1-\frac{\lambda^2}{2} & \lambda & A\lambda^3(\rho-i\eta)\\
-\lambda & 1-\frac{\lambda^2}{2} & A\lambda^2 \\
A\lambda^3(1-\rho-i\eta) & -A\lambda^2 & 1 
\end{array}
\right) + {\cal O}(\lambda^4) \label{eq1}
\ee}

\noi One would like to measure $\lambda$, $A$, $\rho$, and $\eta$ as
precisely and with as much redundancy as possible. In that way, the
unitarity conditions

\be
\sum_i V_{ij} V^\ast_{ik} = \sum_i V^\ast_{ji} V_{ki} = \delta_{jk}
\label{eq2} 
\ee

\noi can be tested. A deviation from expectations in any mode would
signal ``New 
Physics''. Let me discuss some important experiments.

The theoretically cleanest direct measurement of the cornerstone CKM
parameter, $\lambda$, comes from $K_{e3}$ decays $(K\to\pi e\nu)$
\cite{leut} 

\be
\lambda = 0.2196\pm 0.0023 \quad (K_{e3}) \label{eq3}
\ee

\noi where the theoretical and experimental uncertainties (added in
quadrature) are
comparable. That value is to be compared with results from Hyperon and
nuclear beta decays

\bea
\lambda  = & 0.226~\pm0.003 & \quad ({\rm Hyperons}) \label{eq4} \\
\lambda  = & 0.2265\pm 0.0026 & \quad (\beta-{\rm decay}) \label{eq5} 
\eea

\noi There is some inconsistency. Hopefully, ongoing efforts at BNL,
FNAL, and Novosibirsk to
remeasure the $K_{e3}$ decay rates for both $K^+$ and $K_L$ will help  clarify
the situation.

The parameter $A$ is obtained from $V_{cb}$ as measured in
semi-leptonic $B$ decays (the counterpart of $K_{e3}$). Currently, one
finds \cite{ligeti}

\be
A=0.83\pm 0.05 \label{eq6} 
\ee

\noi In Ligeti's talk it was suggested that ongoing and future studies
of $B\to D^\ast e\nu$ decays
may lead to a reduction in the $A$ uncertainty by a factor of 2 or 3 during the
next 3--5 years. Such improvement would be a welcome advancement.

The $\rho$ and $\eta$ parameters are constrained by a combination of
$K$ and $B$ measurements. For example, within the Standard Model, the
CP violating mixing parameter $|\epsilon| = 2.28(1)\times10^{-3}$ provides a
determination of the combination

\be
A^4\lambda^{10}\eta (1-\rho+0.44) \simeq 5.6(1.1)\times10^{-8}
\label{eq7}
\ee

\noi where the error is primarily due to the $K^0$-$\bar K^0$ matrix
element uncertainty. For a given $A$ and $\lambda$, that constraint leads to a
hyperbola in the $\rho$, $\eta$ plane. However, the $\pm20\%$
uncertainty in (\ref{eq7}) is amplified by the current $\pm29\%$ uncertainty
in $A^4\lambda^{10}$. So, the $\rho$, $\eta$ plane is not very favorable for
displaying constraints from $K$ decays. In contrast, it  reduces the
uncertainties in constraints from $B$ 
decays, thus presenting them in a very favorable light. $K$ decay
presentations should resist the lure of the $\rho$, $\eta$ plane.

$B$ physics already  provides some powerful constraints on $\rho$ and $\eta$.
Most useful is the ratio $\Gamma(b\to u)/\Gamma(b\to c)$ which implies
the relatively narrow band \cite{marciano}

\be
(\rho^2+\eta^2)^{1/2} = 0.363\pm0.073 \label{eq8}
\ee

\noi Taken together with the $B^0_d$-$\bar B^0_d$ mixing constraint
$|1-\rho-i\eta | = 1.01\pm 0.22$ and Eq.~(\ref{eq7}), it  suggests (roughly)

\be 
\rho \simeq 0.13 \quad , \quad \eta \simeq 0.34 \label{eq9}
\ee

\noi Other constraints from $B^0_s$-$\bar B^0_s$ mixing and $B\to
J/\psi K_s$ are consistent with values in that general region. Overall,
there is good support for CKM mixing and unitarity.

Given the success of the CKM model, what more can we learn from CP
violation, further CKM studies, and $K$ decays? There are compelling
reasons to push those efforts further. Precision studies of CKM
elements can not only further confirm the standard model, but can help explain
the origin of electroweak mass and perhaps help uncover ``New Physics''.
Indeed, as I will later demonstrate, CP violation and FCNC measurements are
sensitive to effects originating from 
scales as high as 3000 TeV! 

If a true deviation from the Standard Model in $K$ decays or other rare
reactions is uncovered, there will certainly not be a lack of  interesting
explanations. SUSY \cite{hall,mura}, Dynamical Symmetry Breaking, Large Extra
Dimensions, etc.\ can potentially provide new significant sources of CP
violation and FCNC effects.

\section{New CP Violation Results}

The most exciting kaon physics announcement of 1999 was the measurement
of Re$\epe$ by  
$K$TeV \cite{hsiung} and NA48 \cite{sozzi}. Taken together with earlier
studies, those new results 

\bea
{\rm Re} \epe & = & 23.0\pm6.5\times10^{-4} \quad {\rm NA31} \nonumber
\\
& & ~7.4\pm5.9\times10^{-4} \quad {\rm FNAL} \label{eq10} \\
& & 28.0\pm4.1\times10^{-4} \quad K{\rm TeV} \nonumber \\
& & 18.5\pm7.3\times10^{-4} \quad {\rm NA48} \nonumber
\eea

\noi given an average (with PDG expanded error) \cite{sozzi}

\be
({\rm Re}\epe)_{\rm Ave} = 21.2\pm4.6\times10^{-4} \label{eq11}
\ee

\noi That rather solid observation of direct CP violation rules out
(old) Superweak Models. Is it consistent with CKM expectations? Pre
1999, the main theory predictions were (labeled by their home cities)
\cite{bertolini,berttwo}

\bea
({\rm Re}\epe)_{\rm Theory} & = & 4.6\pm3.0\pm0.4\times10^{-4} \quad
({\rm Rome}) \nonumber \\
& = & \left. \begin{array}{c} 3.6\pm3.4\times10^{-4} \\
(10.4\pm8.3)\times 10^{-4}\end{array} \right\} \quad ({\rm Munich})
\label{eq12} \\
& = & 17^{+14}_{-10}\times10^{-4} \quad ({\rm Trieste}) \nonumber 
\eea

\noi The broad range  of those estimates does not allow for a
definitive conclusion. The experimental result does, however, appear to
be somewhat high.

Let me comment on the utility of $\epe$ to probe sources of CP
violation beyond the standard model. It is quite conceivable that some
part of the experimentally observed $\epe$ comes from ``New Physics''.
However, the current theoretical uncertainty of at least $\pm100\%$
(probably more) makes such an interpretation very premature.
Nevertheless, as discussed in Isidori's talk \cite{isidori}, even with
that large a 
theory error one can still obtain interesting constraints on, for
example, potentially large new CP violating $Z_\mu\bar d_L\gamma^\mu s_L$ interactions
induced by SUSY loops in some models \cite{buras}.

The ongoing experiments ($K$TeV and NA48) were, however, designed to
reach a $\Delta \epe$ of $\pm1$--$2\times10^{-4}$, i.e.\  a
$\pm5$--10\% determination of that important quantity. In addition, the
KLOE experiment \cite{anton} at Frascati will provide independent confirmation with
very different systematic uncertainties. It would be a shame if such
elegant measurements could not be fully utilized because of theoretical
shortcomings. 

To significantly reduce the theoretical uncertainty in $\epe$ requires
a systematic first principles calculation of the $K\to2\pi$ amplitudes
in, for example, a lattice gauge theory approach. With today's powerful
QCD teraflop computers and new theoretical methods such as 
domain wall fermions, much more precise calculations may, in fact, be
possible. Indeed, T. Blum \cite{berttwo}described just such an ongoing effort at the
RIKEN BNL Research Center. That collaboration aims for about $\pm20\%$
theoretical uncertainty. Of course, before any new method is accepted,
it must undergo close theoretical scrutiny and pass various consistency
checks. For example, it should quantitatively explain the $\Delta I=1/2$ amplitude
enhancement relative to $\Delta I=3/2$ amplitude in $K$ decays (a
factor of 22). Also,
it should demonstrate control of isospin violating effects which can
feed $\Delta I=1/2$ enhancements into the $\Delta I=3/2$ amplitudes of $\epe$.
Perhaps, most important, as emphasized by Martinelli \cite{berttwo},
the lattice approach should be  self contained. Rather than patch
together pieces of calculations from other prescriptions, it should be
as complete as possible.

If a $\pm20\%$ theoretical calculation of $\epe$ is achieved, it will
provide a very interesting confrontation with experiment. It would
either allow for a powerful precise determination of the Standard Model CP
violation parameter or point to ``New Physics''. Either case justifies
the effort.

Further confirmation of CKM mixing and CP violation is also starting to
come from $B$ decays. (Of course $B$ studies offer tremendous potential
for future studies.) CDF \cite{kroll} has been able to observe an asymmetry in $
\mathop{B}\limits^{(-)}\to J/\psi K_S$. Using a time integrated sample of 400 events, they
have determined (see J. Kroll's talk) $\beta$ of  the unitarity triangle

\be
\sin 2\beta = 0.79^{+0.41}_{-0.44} \label{eq13}
\ee

\noi That result is in good accord with Standard Model expectations
(see M. Gronau's talk \cite{gronau}). Although currently only a $2\sigma$ effect, CDF
expects to reduce the error in Eq.~(\ref{eq13}) by a factor of 5 to
$\pm0.084$. In the longer term, $B$ factories (now up and running),
$B$TeV, and LHC-$B$ hope to achieve $\pm0.02$ precision. The CDF
result indicates that CDF and $D\emptyset$ with their significant upgrades
can be expected to play major roles in future $b$ physics.

Other probes of CP violation discussed at this meeting include:
1) Measurement of $T$ odd asymmetries in $K_L\to\pi^+\pi^- e^+e^-$
and $p\bar p\to K^\pm \pi^\pm\mathop{K}\limits^{(-)}{}^0$, 2) Search for
transverse muon polarization in $K_{\mu3}$ decay, 3) Hyperon decay
asymmetries, and 4) Electric dipole moments.

The measured experimetal 13.6\% $T$ odd asymmetry between the
$\pi^+\pi^-$ and $e^+e^-$ planes in $K_L\to\pi^+\pi^-e^+e^-$ observed
at Fermilab (see Ladovsky's talk \cite{ledov}) is in good accord with the 13--14\%
expectation due to $\epsilon$ in the Standard Model (see talks by
Sehgal \cite{sehgal} and Savage \cite{savage}). That result was based on 1811 events at KTeV\null.
Such a large asymmetry in $K$ decays is quite spectacular and was to
most people very
surprising. Future efforts at KAMI could yield $10^5$ decays in that
channel and perhaps provide another probe of direct CP violation.

We were also reminded here of an earlier $T$ odd study from CP LEAR
\cite{bloch} 

\be
A= \frac{R(\bar K^0\to\pi^-e^+\nu) - R(K^0\to \pi^+e^-\bar \nu)}{R(\bar
K^0 \to\pi^-e^+\nu) + R(K^0\to\pi^+e^-\bar\nu)} = 6.6\pm 1.3\pm
1.6\times10^{-3} \label{eq14}
\ee

\noi That Kabir test is in good agreement with the Standard Model
prediction $A=6.4\times10^{-3}$, again  due to $\epsilon$.

G.-Y. Lim reported a recent KEK result for the muon transverse
polarization in $K_{\mu3} (K^+\to\pi^0\mu^+\nu_\mu)$ decay \cite{lim}

\be
p^T_\mu = \hat s_\mu \cdot (\hat p_\mu\times\hat p_\pi) \label{eq15}
\ee

\noi They have reached

\be
p^T_\mu = -0.0042\pm 0.0049\pm0.0009 \label{eq16}
\ee

\noi and aim for $10^{-3}$ sensitivity. The Standard Model predicts
$P^T_\mu\sim 0$; so, a non-zero experimental result would directly point to a
new source of CP violation. The leading candidate would be a charged
Higgs exchange amplitude with a relatively large CP violating phase \cite{peccei}.
Such direct searches for completely new sources of CP violation are
extremely important and must be pushed as far as possible. An
approved BNL experiment would reach $10^{-4}$ sensitivity, but
unfortunately, it may never get to take data because of uncertainties in
future AGS running for fixed target experiments.

Larger than expected CP violating asymmetries in Hyperon decays (see
talks by Pakvasa, White, and Solomey \cite{pakvasa}) could also point to ``New
Physics''. An extensive Hyperon decay program is being proposed at
Fermilab.

Perhaps the most promising way to uncover new sources of CP violation
is the study of electric dipole moments. Such effects are predicted to
be non-zero, but unobservably small in the CKM framework. However,
``New Physics'' of the type needed in some Baryogenesis scenarios
\cite{worah} , for
example, could provide much larger edm signals, near the current
experimental bounds. In the talk by M. Romalis \cite{romalis} we heard
of ambitious 
efforts to push the sensitivity for the neutron and electron edm's from
$6.3\times10^{-26}\to 10^{-28} e$-cm and
$4\times10^{-27}\to10^{-31}e$-cm respectively. A proposal by the
$g_\mu$-2 collaboration at BNL would also greatly extend the search for
a muon edm from $10^{-18}\to10^{-24}e$-cm. All such advances should be
strongly encouraged, since a positive finding would be revolutionary
and may, in fact, be just waiting to be unveiled.

A general theoretical framework for discussing CPT and Lorentz
invariance violation was given by A. Kostelecky \cite{kostel}. $K$ physics studies
currently provide the most sensitive tests of CPT \cite{difalco}. Measurements of
$m_{K^0}$-$m_{\bar K^0}$ at $K$TeV, NA48, and CP LEAR have reached the
incredible $10^{-18}$ GeV level and are beginning to approach the
interesting $m^2_K/m_{\rm planck}$ sensitivity. Future measurements at
Frascati (see S. DiFalco's talk on KLOE) will further advance the cause.

\section{Medium Rare, Rare, and Forbidden Decays}

In recent years, great progress has been made in the study of flavor
changing neutral current (FCNC) decays of $K$ mesons. Experimental
studies have been accompanied by an expansion in our theoretical
arsenal of tools which now includes: chiral perturbation theory, large
$N_c$, lattice gauge theories etc. Together, they have allowed us to
test the Standard Model as well as to probe for and constrain possible
``New Physics''. Here, I divide rare decays into three categories: 1)
Medium Rare which includes roughly $10^{-5}$-$10^{-9}$ branching
ratios, 2) Rare decays with branching ratios $\lsim 10^{-9}$ and 3)
Forbidden decays which do not occur in the Standard Model. For the last
of those, I will discuss only muon-number non-conservation, because it
provides such a sensitive probe of ``New Physics''.

At this meeting, we heard about many measurements of medium rare
decays \cite{zeller}. In table~\ref{tab2}, I list some of the results that were
discussed. Most impressive to me is the fact that some measurements of
historical importance such as $K_L\to\mu^+\mu^-$ have gone from a
handful of events to precision measurements based on 5--10 thousand
events \cite{molzon}. Indeed, they now confront the standard model at its quantum
loop level so as to constrain ``New Physics'' such as SUSY or
Technicolor inspired models. In addition, the abundance of events in
$K_L\to\mu^+\mu^-$, $\pi^+\pi^-e^+e^-$, $K^+\to\pi^+\mu^+\mu^-$ etc.\
suggest that they may be further used to study CP violation effects in the
future. Note, also that those measurements have been very useful in
fine tuning the parameters of chiral perturbation theory and advancing
its techniques (see talk by J. Bijnens \cite{bijnens}).

\begin{table}[htb]
\begin{center}
\caption{Examples of Medium Rare $K$ Decay Branching Ratios
\label{tab2}}
\begin{tabular}{lll}
Decay Mode \qquad & Branching Ratio & Comments \\ \\
$K^+\to\pi^+e^+e^-$ & $2.82\pm0.04\pm0.07\times10^{-7}$ & BNL E865 -
Preliminary \\
$K^+\to\pi^+\mu^+\mu^-$ & $9.23\pm0.6\pm0.6\times10^{-8}$ & {\tt   "
"} \\
$K^+\to \pi^+\pi^-e^+\nu_e$  & $\sim3.9\times10^{-5}$ & \qquad (300,000
events) \\
$K^+\to\pi^+\pi^0\gamma$ & $4.72\pm0.77\times 10^{-6}$ & BNL E787 \\
$K_L\to\pi^+\pi^-e^+e^-$ & $4.4\pm1.3\pm0.5\times10^{-7}$ & KEK E162 \\
$K_L\to e^+e^-\gamma$ & $1.06\pm0.02\pm0.02\pm0.04\times10^{-5}$ & CERN
NA48 \\
$K_L\to \mu^+\mu^-$ & $7.18\pm0.17\times10^{-9}$ & BNL E871 
\end{tabular}
\end{center}
\end{table}

Rare $K$ decay experiments have made spectacular progress in measuring incredibly
small branching ratios or pushing bounds. D. Ambrose \cite{ambrose} reported the
smallest branching ratio ever measured in a decay process

\be
B(K_L\to e^+e^-) = 8.7^{+3.7}_{-4.1}\times10^{-12} \label{eq17}
\ee

That result is in good agreement with the Standard Model prediction of
$9\times10^{-12}$. It indicates that even such rare decays can be cleanly
observed and measured with precision. That bodes well for other more
interesting rare decays for which only bounds currently exist \cite{whit}

\bea
B(K_L\to\pi^0e^+e^-) & < & 5.6\times10^{-10} \qquad\qquad\qquad\qquad
(a) \nonumber \\
B(K_L\to\pi^0\mu^+\mu^-) & < & 3.4\times10^{-10}
\qquad\qquad\qquad\qquad (b) \label{eq18} \\
B(K_L\to\pi^0\nu\bar \nu) & < & 5.9\times10^{-7}
\qquad\qquad\qquad\qquad\;\, (c) \nonumber 
\eea

\noi but are expected to occur at about $5\times10^{-12}$,
$1\times10^{-12}$ and $3\times10^{-11}$ respectively. Each of those decays
provides a nice test of direct CP violation, if a real measurement can
be achieved. The golden mode \cite{litten} $K_L\to\pi^0\nu\bar\nu$ is particularly
attractive because it is theoretically pristine (with only about
$\pm1$--2\% theoretical uncertainty). In fact, as I will subsequently
describe, it has the unique potential of determining the extremely
important Jarlskog \cite{jarls} CP violating parameter $J_{CP}$ at
about the 
$\pm5\%$ level. Such a measurement is so compelling, that it must be
carried out, if experimentally feasible (more commentary and discussion
of experimental goals later).

Similar to $K_L\to\pi^0\nu\bar\nu$ is the rare decay $K^+\to
\pi^+\nu\bar\nu$ being pursued by the E787 collaboration at BNL\null.
That group saw a single event in its 1995 run. Further analysis, as
described by G. Redlinger \cite{redl}, did not uncover additional candidates. The
collaboration has not updated their 1 event branching ratio, but one
expects that it now corresponds to about $1.5\times10^{-10}$ with
fairly large errors. About 2 times as much data remains to be analyzed,
but already the experiment appears to be consistent with the Standard
Model expectation $B(K^+\to\pi^+\nu\bar\nu)\simeq 0.9\times10^{-10}$. 

The theoretical error \cite{buch} on $B(K^+\to\pi^+\nu\bar\nu)$ due to charm mass
and QCD uncertainties is only about $\pm7\%$. So, it would be extremely
useful to measure that branching ratio with a similar $\pm10\%$
experimental error,  both as a means of determining
CKM mixing parameters and constraining ``New Physics''. E787 could wind
up with several events when the analysis is complete. Its approved
follow-up E949 at BNL has a goal of $0.8\times10^{-11}$ sensitivity, or
about 10 Standard Model events (about a $\pm30\%$ determination of
$B(K^+\to\pi^+\nu\bar\nu)$). In the longer term, the CKM proposal at
Fermilab's KAMI facility would aim for 100 events or $\pm10\%$. As I
will describe later, a $\pm10\%$ measurement of that important branching
ratio will allow ``New Physics'' to be probed beyond the 1000 TeV (PeV)
level!

Muon-number violating (forbidden) decays have also been searched for
with impressive sensitivities. Kaon and muon decays have achieved the
bounds \cite{molzon} given in Table~\ref{tab3}. If no events appear in the ongoing
E865 analysis at BNL, the bound on $K^+\to\pi^+\mu e$ is expected to
reach $8\times10^{-12}$. Searches for those forbidden $K$ decays could
probably be pushed by about another order of magnitude at future high
intensity kaon facilities. However, currently, most planning activity
involves forbidden muon decays such as $\mu^+\to e^+\gamma$ and
$\mu^-N\to e^-N$ (coherent muon conversion in muonic atoms) because
ideas for extending the current experimental sensitivity by 3 or 4
orders of magnitude exist. (New forbidden decay searches should
generally strive for at least 2 orders of magnitude improvement.)

\begin{table}[htb]
\caption{Current bounds on muon-number violating decays and future
potential. \label{tab3}}
\medskip
\hbox{ Decay~Mode\qquad\qquad  Current~Bound \qquad\qquad\qquad \, 
Future~Potential\hss}
\[
 \left.
\begin{array}{ll}
B(K_L\to\mu e) & \quad<4.7 \times 10^{-12} {\rm~~BNL~E781 } \\
B(K^+\to\pi^+\mu e) & \quad<4.8\times10^{-11} {\rm~~BNL~E865} \\
B(K_L\to\pi^0\mu e) & \quad<3.2\times10^{-9} {\rm~~FNAL-KTeV} 
\end{array} \right\}  {{\rm Probably~could~be~pushed} \atop 
{\rm to~a~few}\times 10^{-13}} \]
\begin{center}
\begin{tabular}{lll}
$B(\mu^+\to e^+\gamma)$ & $\,<1.2\times10^{-11}$
MEGA & \quad$10^{-14}$  PSI Proposals \\
$B(\mu^+\to e^+e^-e^+)$ & $\,<1\times10^{-12}$ &
\qquad\qquad --- \\
$B(\mu^-N\to e^-N)$ & $\,<6\times10^{-13}$
~~SINDRUM II & $\quad\,5\times10^{-17}$ MECO at BNL 
\end{tabular}
\end{center}
\end{table}

Coherent muon-electron conversion, $\mu^-N\to e^-N$, is a particularly
powerful probe of ``New Physics''. Its discovery potential is very
robust, including SUSY loops, heavy neutrino mixing, $Z^\prime$ bosons,
Multi-Higgs models, compositeness etc. To demonstrate its reach,
consider the muon number non-conserving four fermion interaction

\be
{\cal L} = \frac{4\pi}{\Lambda^2} \eta_q\bar e\gamma_\alpha \mu \bar
q\gamma^\alpha q \qquad q=u,d \label{eq19}
\ee

\noi where $\Lambda$ is a generic scale of ``New Physics'' and $\eta_q$
represents a model dependent combination of couplings, mixing
parameters, etc. At a sensitivity of $5\times10^{-17}$, the goal of the
proposed MECO experiment at BNL, one is probing (approximately)

\be
\Lambda \gsim 3000{\rm~TeV} \, \sqrt{\eta_q} \label{eq20}
\ee

\noi Few experiments are capable of exploring such short-distance
scales. Of course, a discovery would be revolutionary. Given its
potential, experiments such as MECO must be pushed as far and as soon
as possible. 

\section{Quantitative Tests of CKM Unitarity - CP Violation}

The $3\times3$ CKM mixing matrix, $V_{CKM}$, must be unitary. A
convenient parametrization

\be
V_{CKM} = \left(\begin{array}{ccc}
c_1c_3 & s_1c_3 & s_3 e^{-i\delta} \\
-s_1c_2-c_1s_2s_3e^{i\delta} & c_1c_2-s_1s_2s_3e^{i\delta} & s_2c_3 \\
s_1s_2-c_1c_2s_3e^{i\delta} & -c_1s_2-s_1c_2s_3e^{i\delta} & c_2c_3
\end{array} \right) \quad \begin{array}{c}
 \\ c_i=\cos\theta_i \\ s_i=\sin\theta_i \end{array} \label{eq21}
\ee

\noi exhibits the features that allow the orthonormal relationships in
Eq.~(\ref{eq2}) to be satisfied.

One can test the Standard Model and search for ``New Physics'' by
making clean precision measurements of $V_{CKM}$ elements and seeing
if unitarity is satisfied. For example, 4 measurements determine
$\theta_1$, $\theta_2$, $\theta_3$ and $\delta$ (or Wolfenstein's
$\lambda$, $A$, $\rho$, and $\eta$). A fifth measurement then tests
unitarity. Alternatively, each of the individual relationships in
Eq.~(\ref{eq2}) can be tested by 3 (or more) measurements. Unitarity can be
tested within $K$ or $B$ decays alone or in comparison with one
another. Of course, in all cases theoretical uncertainties should be
minimized. Also, it is useful to have as many different consistency checks as
possible, since that allows many potential ``New Physics'' effects to
be explored.

CP violation and FCNC effects are particularly good probes of ``New
Physics'', because the Standard Model predictions are generally so small. Which
system is more sensitive, $K$ or $B$ decays? How does one compare the
potential of $K$
and $B$ studies in an unbiased manner? A nice answer is provided by
Cecilia Jarlskog's $J_{CP}$ parameter. Let me describe its utility.

The six orthogonal relations in Eq.~(\ref{eq2}) with $j\ne k$ give rise
to so-called unitarity triangles. I will label the 6 distinct triangles
by their $(j,k)$ indices. The (1,3) or $(d,b)$ triangle

\be
V_{ud}V^\ast_{ub} + V_{cd}V^\ast_{cb} + V_{td}V^\ast_{tb} = 0
\label{eq22}
\ee

\noi is best known because of its general use in illustrating $b$ physics studies. $B$
programs aim to measure the angles and sides of those triangles in as
many ways as possible. A deviation from closure or single inconsistent
measurement 
would signal ``New Physics''. In addition, if one factors out
$V_{cd}V^\ast_{cb}$ from that relation, the remaining triangle is
nicely illustrated in the $\rho$, $\eta$ plane.

In $K$ physics there is also a useful unitarity triangle, the (1,2) or
$(d,s)$ relation

\be
V_{ud}V^\ast_{us} + V_{cd}V^\ast_{cs}+V_{td}V^\ast_{ts} = 0
\label{eq23}
\ee

\begin{figure}[h]
\hspace*{15mm}
$B$ Physics: 
$V_{ud}{V_{ub}}^* + V_{cd}{V_{cb}}^* + V_{td}{V_{tb}}^*  = 0 $

\vspace*{7mm}

\hspace*{40mm}\psfig{figure=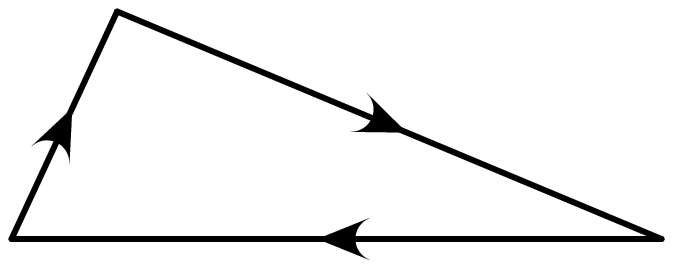,width=36mm}

\begin{picture}(0,0)(0,0)
\put(160,48)  {$V_{td}{V_{tb}}^*$}
\put(70,30)  {$V_{ud}{V_{ub}}^*$}
\put(150,-3)  {$V_{cd}{V_{cb}}^*$}
\end{picture}

\vspace*{2mm}
\hspace*{35mm}$J_{13}^{\rm CP} = 2\times \mbox{Area}$

\vspace*{12mm}

\hspace*{15mm}
$K$ Physics: 
$V_{ud}{V_{us}}^* + V_{cd}{V_{cs}}^* + V_{td}{V_{ts}}^*  = 0 $

\vspace*{7mm}
\hspace*{-14mm}
\psfig{figure=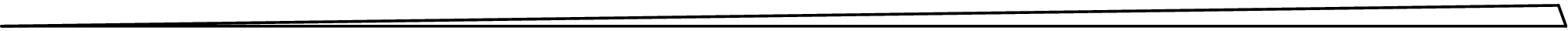,width=145mm}

\begin{picture}(0,0)(0,0)
\put(380,15)  {$V_{td}{V_{ts}}^*$}
\put(180,-3)  {$V_{ud}{V_{us}}^*$}
\put(180,25)  {$V_{cd}{V_{cs}}^*$}
\end{picture}

\vspace*{5mm}

\hspace*{15mm}
$J_{12}^{\rm CP} = 2\times \mbox{Area} = 5.60\left[ {\rm BR}
(K_L\to \pi^0\nu\bar \nu) \right]^{1/2}$

\vspace*{5mm}
\caption{Unitarity triangles for $B$ and $K$ studies. \label{fig1}}
\end{figure}

\noi Both triangles are illustrated in fig.~\ref{fig1}. The (1,2)
triangle has angles near 0 and $90^\circ$ which imply very small CP
violating decay asymmetries (in contrast with $B$ decays). Does that
make it uninteresting? No. As pointed out by C. Jarlskog, the most
interesting feature of any unitarity triangle is its area and that
quantity is the same for all 6 triangles.

\begin{quote}
``All CKM triangles are created equal in Area!''
\end{quote}

In fact, she observed that a quantity $J_{CP} = 2\times{}$the triangle
area was the unique real measure of CP violation in the Standard Model.
Unitarity requires 

\be
J_{CP} = J_{12} = J_{13} = J_{23} = J_{21} = J_{31} = J_{32}
\label{eq24}
\ee

\noi In terms of the parametrizations of Eq.~(\ref{eq21}) or
Eq.~(\ref{eq1})

\be
J_{CP} = s_1s_2s_3c_1c_2c_3^2\sin\delta \simeq A^2\lambda^6\eta
\label{eq25}
\ee

\noi Standard model CP violation is tested by measuring $J_{CP}$ as
precisely and in as many distint ways as possible \cite{buchtwo}. A
deviation would signal ``New Physics''

Currently, a global fit to all $K$ and $B$ studies indicates \cite{buras}

\be
J_{CP} = 2.7\pm1.1\times10^{-5}, \label{eq26}
\ee

\noi i.e.\ it is determined to about $\pm40\%$. How well can the next
generation of $K$ and $B$ studies individually determine $J_{CP}$? In
the case of $B$ physics, the long term prospects are that $J_{13}$
will be measured to about $\pm15\%$. Pushing to $\pm5\%$ is extremely
difficult, but worth trying to achieve.

In the case of $K$ decays, we are extremely fortunate. The decay
$K_L\to\pi^0\nu\bar\nu$ directly determines the height of the (1,2) triangle and
the base is already well known from $\beta$-decay and $K_{e3}$ decays. One finds

\be
J_{12} = J_{CP} = 5.60 [ B(K_L\to\pi^0\nu\bar\nu)]^{1/2} \label{eq27}
\ee

\noi That result is extremely clean. Theoretical uncertainties are at
the level of 1--2\%. So, the only real limitation is how well
$B(K_L\to\pi^0\nu\bar\nu)$ can be measured. A proposed measurement at
the $\pm25\%$ level (about 16 events) would determine $J_{CP}$ to
about $\pm12$ 1/2\%, which is better than long term $B$ physics expectations. In
the longer term, a 10\% measurement of $B(K_L\to\pi^0\nu\bar\nu)$ would
give $J_{CP}$ to $\pm5\%$. Of course, we need at least 2 measurements
of similar precision $J_{CP}$ for comparison; so, it would be nice if
$B$ efforts could remain competitive.

How might determinations and comparison  of $J_{12}$ and $J_{13}$ with high precision
be utilized? As a simple illustration, consider a strangeness changing
interaction \cite{marctwo}

\be
{\cal L} = \frac{4\pi}{\Lambda^2} B\bar d_L \gamma_\alpha s_L \bar
\nu_i \gamma^\alpha\nu_i + h.c. \label{eq28}
\ee

\noi do to ``New Physics'' at scale $\Lambda$. A $\pm10\%$ measurement
of $B(K_L\to\pi^0\nu\bar\nu)$ would probe $\Lambda\sim3000$ TeV $({\rm
Im}B)^{1/2}$. Note that $\pm10\%$ precision in $B(K^+\to
\pi^+\nu\bar\nu)$ provides similar probing power. Clearly, studies of
$K^+\to\pi^+\nu\bar\nu$ and $K_L\to\pi^0\nu\bar\nu$ must be pushed as
far as possible.

\section{Concluding Remarks (Future Outlook)}

Direct CP violation in $K\to2\pi$ decays has finally been unambiguously
observed. Ongoing 
experimental efforts should eventually determine Re$\epe$ to $\pm5$--10\%.
Theoretical calculations must strive to reach a similar level of
precision.

$B$ physics has come of age. Studies at CLEO will soon share the
spotlight and  be challenged by
asymmetric $B$ factories with CP violation as their primary goal. CDF
and $D\emptyset$ will also  be important players in the future along
with LHCB, TeV$B$ etc.

$B$ studies open a new exciting frontier, but they do not close the
door on $K$ or rare muon decays. The Kaon system is still the best place to
look for CPT violation and the new $\phi$ factory at Frascati will be
at the forefront of that effort. The rare decays
$K^+\to\pi^+\nu\bar\nu$ and $K_L\to\pi^0\nu\bar\nu$ are exceptionally clean
theoretically. Besides testing CKM mixing with great precision, they
are capable of probing ``New Physics'' up to about the 3000 TeV level.
The muon 
number violating reaction $\mu^-N\to e^-N$, similarly probes 3000 TeV
physics, but in a very different channel. Such outstanding experimental
opportunities are extremely scarce. They must be seized and  pushed as far as
possible.

The decay $K_L\to\pi^0\nu\bar\nu$ is very special. It alone can determine the
all important Jarlskog parameter $J_{CP}$ to about $\pm5\%$ in the long
term. It would then set the standard for comparing other manifestations
of CP violation in $K$ and $B$ decays. It must be pursued with the same
zeal and priority as $B$ physics.

Other rare $K$ decays, $K_L\to\pi^0 e^+e^-$, $K_L\to\pi^0\mu^+\mu^-$,
$K^+\to\pi^+\mu^+\mu^-$ etc.\ can also contribute to our understanding
of CP violation and search for ``New Physics''. Kinematic and
polarization asymmetries may be particularly useful in those endeavors. 

Kaon physics has had a glorious history. It continues to be exciting
(e.g.\ $\epe$, $K^+\to\pi^+\nu\bar\nu$, $K_L\to\pi^0\nu\bar\nu$ etc.)
Are there any future big surprises or great discoveries waiting
still to be uncovered in the kaon system? We will find out only if we
continue to expand our efforts and follow our instinct to explore.

\end{document}